# The Emerging Web of Social Machines

Silvio R. L. **Meira**[1,3], Vanilson A. A. **Buregio**[1], Leandro M. **Nascimento**[1], Elaine G. M. de **Figueiredo**[1], Misael **Neto**[2,3], Bruno P. **Encarnação**[3], Vinícius **Garcia**[1,3]

[1]Federal University of Pernambuco - UFPE, Informatics Center, Recife, PE, Brazil: www.cin.ufpe.br

[2]Catholic University of Pernambuco – UNICAP, Recife, PE, Brazil: www.unicap.br

[3]Recife Center for Advanced Studies and Systems - C.E.S.A.R, Recife, PE, Brazil: www.cesar.org.br

Corresponding author: silvio@meira.com, twitter.com/srlm

## Abstract

We define a notion of social machine and envisage an algebra that can describe networks of such. To start with, social machines are defined as tuples of input, output, processes, constraints, state, requests and responses; apart from defining the machines themselves, the algebra defines a set of connectors and conditionals that can be used to describe the interactions between any number of machines in a multitude of ways, as a means to represent real machines interacting in the real web, such as Twitter, Twitter running on top of Amazon AWS, mashups built using Twitter and, obviously, other social machines. This work is not a theoretical paper as yet; but, in more than one sense, we think we have found a way to describe web based information systems and are starting to work on what could be a practical way of dealing with the complexity of this emerging web of social machines that is all around us. *This version should be read as work in progress and comments, observations, bugs... are most welcome and should be sent to the email of the first, corresponding author.*

## Disclaimer

**This is a very early report of work in progress**, being made available in this version and timing to help creating context for the submission of proposals for the MSc and PhD degrees at Federal University of Pernambuco at Recife. The authors welcome all sorts of comments and contributions; at some stage in the future, we hope to submit a later version of this report to be considered for publication in a learned journal.

## Acknowledgment

This work was partially supported by the **National Institute of Science and Technology for Software Engineering (INES**, www.ines.org.br), funded by CNPq and FACEPE, grants 573964/2008-4 and APQ-1037-1.03/08.

# 1. Introduction

The traditional concept of software has been changing during the last decades. Since the first definition of a computing machine described by [Turing, 1937], software started to become part of our lives and has been turned pervasive and ubiquitous with the introduction of personal computers, the internet, smartphones and, of later, the internet of things. In fact, one can say that software and the internet changed the way we communicate, the way business is done and the internet is changing the way software is developed, deployed and used. Nowadays, computing means connecting [Roush, 2005]; and it just may be the case that developing software is the same as connecting services.

The early internet was a web of mostly static content, basically HTML pages presented in a read only mode or else systems with a very simple transactional capability from the user's point of view (think a search engine's interaction "box"); this is the web we could classify as "1.0". As a further development, simultaneous with the appearance of new technologies and the notion of Ajax [Garrett, 2005], web pages became more interactive and allowed content sharing, social interaction and collaboration, which led to blogs, wikis and social networks, we had the read/write web, which is also known as web "2.0". From then on, we all can clearly see that a new phase is emerging, the web "3.0", the web as a programming platform, the network as an infrastructure for innovation, on top of which all and sundry can start developing, deploying and providing information services using the computing, communication and control infrastructures in a way fairly similar to utilities such as electricity.

The web 3.0 is the networked space-time where innovation lies on the power of developing software for the web, through the web, and in the web, using the web as both programming platform (in lieu of the usual computer/operating system/development environment platform) and deployment and execution environment. Several examples of this (let's say) scenario are current developments in Facebook, Twitter, Yahoo!, Salesforce, Google, Amazon and many other corporations are making their APIs available for anyone to develop applications that interact with their services.

Although there have been many studies about the future of the internet and concepts such as web 3.0, programmable web [Yu & Woodard 2009, Hwang et al. 2009], linked data [Bizer et al. 2009, Halb et al. 2008] and semantic web [Hitzler et al. 2009], the segmentation of data and the issues regarding the communication among systems obfuscates the interpretation of this future. Kevin Kelly, of Wired fame, is quoted as having said once: "*The internet is the most reliable machine ever made. It's made from imperfect, unreliable parts, connected together, to make the most reliable thing we have*". Unstructured data, unreliable parts and problematic, non-scalable protocols are all native characteristics of the internet that has been evolving for 40 years; at the same time, they are the good, the bad and the ugly of a web in which we rely more and more in the everyday life of everything, that needs a unifying view and explanations in order to be developed, deployed and used in a more efficient and effective way.

Indeed, the web is changing in a fundamental way and approaches such as SOA, REST, XaaS, Cloud Computing each play important roles in this emerging web. However, the read/write and programmable webs are recent enough to represent very serious difficulties in understanding their basic elements and how they can be efficiently combined to develop real, practical systems in either personal, social or enterprise contexts. There has not been a clear, precise description of each and every entity on this new emerging web (above the basic, 1.0, which is a restriction of it) and we believe it is necessary to create new mental models of such a web as a platform, in order to provide a common and coherent conceptual basis for the understanding of this young, upcoming and possibly highly innovative phase of software development.

This paper tries to explain the web in terms of a new concept named Social Machines (SM). This is not a theoretical paper as yet; but, in more than one sense, we think that this work can collaborate to the process of providing a unifying vision to describe web based information systems and are starting to work on what could be a practical way of dealing with the complexity of this emerging web of social machines.

The remainder of this paper is organized as follows. Section 2 discusses some related work. Section 3 presents the concept of Social Machines and sets up their main elements, characteristics and types. Section 4 shows the outcomes obtained from the application of this concept in a practical case study and, finally, Section 5 presents some concluding remarks and directions for future work.

## 2. Related Work

Although the concept of Social Machines overlaps other research fields and issues currently well studied such as SaaS, Cloud Computing, SOA and Social Networks, we have not found any research that deals with the concept as we do propose herein. Some authors had already mentioned the term Social Machines, as [Roush, 2005]. However, the expression has been used with a different meaning, representing human operated machines responsible for socializing information among communities, that is, an intersection of the areas and studies of social behavior and computational systems.

The notion of social machines described in this work is not related to or does originate from Deleuze and Guattari's; theirs are virtual machines operating in given social fields [Patton, 2000]. It is nonetheless interesting that they (or indeed Guattari in [Guattari, 1995]) thought of language and mass media as "large scale social machines" (as discussed in [Fuglsang & Sørensen, 2006]), systems that (in society) consume, produce and record (information) and, maybe even more relevant to our approach, are connected at large.

In contemporary robotics, there is a notion of social machine, but that is also unrelated to what is proposed here; the robotics view of a social machine is that of one that can relate to people, i. e., can demonstrate empathy to humans. But attempts to build "empathy machines" [Eller & Touponce 2004] are not new: the japanese *chahakobi ningyo*, windup dolls that are capable of serving tea [Hornyak, 2006] are centuries old and antecedents of japanese robots of all sorts that, somehow, interact with humans nowadays [Kroker & Kroker, 2008].

Having said so, it is very likely that the notion of social machines introduced here was waiting to be described for some time now, reason why we do not claim to have invented it but rather to have discovered a new interpretation of the expression in a particular setting, that of a network of programmable machines that are connected to each other and also connect people and institutions in a web of computing, communication and control that needed a much more abstract description and formalization than its external behavior in the form of a public (web) interface and number of APIs on top of the *de facto* standard internet protocols.

In what follows, we are going to try to define the web of programmable machines in terms of the interpretation we are assigning to the expression "social machines", without any hope of having covered the whole of the subject, its wider implications and, even more important, its foundational theories.

## 3. Social Machines: the Concept

In general, a Social Machine (SM) represents a connectable entity containing an internal *processing unit* and a wrapper interface that waits for *requests* from and replies (with *responses*) to other social machines. Its processing unit receives *inputs*, produces *outputs* and has *states;* and its *connections* define intermittent or permanent relationships with other SMs, connections which are established under specific sets of *constraints*. Figure 1 illustrates a basic representation of a Social Machine.

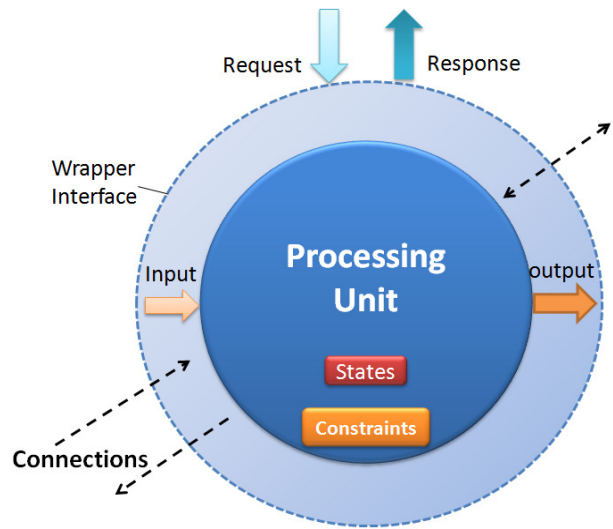

Figure 1: A representation of a Social Machine.

We define a SM as a tuple of *Relationships, Wrapper Interface, Request, Response, State, Constraints, Input, Processing Unit* and *Output*, as following:

SM = <Rel, WI, Req, Resp, S, Const, I, P, O>

We are still doing the basic research on an algebra that can be used to describe social machines, their networks and interactions between any number of them; for now, let us define each element of a SM as:

- **Relationships**, <Rel>: A SM can connect with other SMs following any well defined protocol. The concept of relationships between SMs is similar to that of relationships between people; we can view them as trusted relations between different SMs, satisfying established constraints. For example, if a specific SM intends to establish a trust relationship with Twitter.com it needs to meet twitter's connection contraints[1]. This gives us the liberty to connect any number of SMs through the Web in order to form different networks and implement new services from the ones that already exist. The types and degrees of these relationships are explained in more details in Section 3.2.

- **Wrapper Interface**, <WI>: wrapper interface is a communication layer through which a SM externalizes its services and allows interactions with other SMs on the web. For example, considering Twitter as a SM, the API it provides can be considered a wrapper interface. Through Twitter's API it is possible to interact with its main services (search, tweet, direct messaging, retweet...). It is important to highlight that APIs represent one instance of interactions. Facebook, for example, uses email notification to alert users of new wall posts; in this case, Facebook's email service is a another type of wapper interface.

- **Request**, <Req>: A request can be seen as a remote procedure call to the services provided by an SM's wrapper interface. Requests can be of two types:

---

[1] http://dev.twitter.com/pages/rate-limiting

Paper still in α; new versions will be announced at twitter.com/srlm, #SocMac. 4

- i. *SM's functionality request* (F-Req): A *F-Req* is an invocation of the core functionalities implemented by the SM's processing unit. Depending on the signature of the considered functionality, the request may or not contain input parameters.
- ii. *SM's meta-information request* (MI-Req): A *MI-Req* can be seen as an invocation which queries a given Social Machine about itself. These requests are useful for retrieving meta-information associated to an SM such as cost and response time of its services; required parameters and their data types; possible and current status; usage policy and existing constraints, etc.

- **Response**, <Resp>: A response can be seen as a remote reply to other SM's through the wrapper interface. Responses can be of three types:
  - i. *SM's functionality response (*F-Resp): A F-Resp is a direct reply to an SM's functionality request (F-Req). A sorting SM may receive an array and a sorting method as parameters; if so, this SM will process the parameters and respond to the request with the sorted array. The relationship between F-Req and F-Resp is of kind 1:{0..n}, that is, for any given request there might be none, one or many responses. For example, a given SM A can make a request to another SM B, asking B to produce a response every time its status changes. This can naturally make B give back none, one or infinitely many responses to A.
  - ii. *SM's meta-information response (*MI-Resp*):* MI-Resp is a direct reply to SM's meta-information request MI-Req.
  - iii. *Notification response (*N-Resp*):* these responses notify/alert connected SMs about different conditions of a SM, possibly carrying exceptional data, informing the requester of the occurrence of runtime errors, bad parameter errors, constraint violations, success messages, etc.

- **State**, <S>: A SM may or may not maintain its current state(s). For example, a very simple stateless SM, Succ could, once given a natural number, return its successor, keeping no track of what it is doing. On the other hand, if a SM maintains its state, it may provide a way to access such information through an MI-Req. For example, if we consider Twitter [Twitter, 2010] as a SM, any request asking for its state can get a response informing something like "fully operational" or "over capacity".

- **Constraints**, <Const>: Any restrictions that a given SM can have are described here. Constraints can be directly compared to non-functional requirements in the software life cycle. For instance, if we consider a web server as a SM, one of its constraints could be the maximum number of concurrent accesses that would lead to a denial of service error. Hence, constraints can be used as rules to be considered during the establishment of relationships among different SMs. They can specify, for instance, authorization protocols (for security), number of requests per hour (for performance) and additional properties which can influence other quality attributes.

- **Input**, <I>: Data handled by a SM's processing unit, exactly as the input (parameter) of a function in any programming language. For example, as mentioned in the Succ SM, the input could be the number 2 and the output for this input would be the number 3.

- **Processing unit**, <P>: This element represents a process, algorithm or the combination of existing SMs that provides the core functionalities of a Social Machine. In other words, this represents the internal computational unit implemented by the SM in order to accomplish the services it is intended to provide.
- **Output**, <O>: After processing any given input, a SM may return a result; of course it is possible for a SM to return no output for a given input.

Other examples of Social Machines are shown in Section 4, where we explain a practical case study.

## 3.1. Main Characteristics

After presenting the schematic concept of a SM, we are now able to highlight some of its main characteristics, as following:

- **Sociability**: By the very nature of the concept we are proposing, SMs are sociable stuff and, in nearly all cases, each one should provide means to interact with one another. The isolated, autistic social machine is an exception. SMs is not only a possible foundation for describing the sharing and reuse of networked information systems but should be implemented in a way to provide for that. The idea behind Social Machines is to take advantage of the networked environment they are in there in to make ir easier to combine and reuse exiting services from different SMs and use them to implement new ones.
- **Compositionality**: Apart from the basic "combinator" SMs, any higher complexity SM can be represented in terms of other SMs having lower complexity and/or fewer "social skills". This allows for the description and implementation of SMs using a "divide to conquer" approach, an example of which is presented in the case study of Section 3.2.
- **Platform independency**: Social Machines are platform, technology and implementation independent; at the highest level of abstraction, we can think of the approach proposed herein as a novel way to describe the architecture of (web based, web intensive) information systems.
- **Implementation independency**: A SM should provide its services in a way that other SMs making use of such services do not have to care about how they were implemented. Besides, it is desirable that SMs use well defined and de facto standard protocols so that the communication between them is as simples as possible. The SMs' Wrapper Interface should be legible and easy to use; the clearer a Wrapper Interface is in providing access to its services, the easier will be is for others to use their services.
- **Self-awareness**: Every social machine (with the exception of the exceptional, "autistic" ones) should be self-aware, meaning that it must be able to answer the request "Who are you?" with an answer such like: "I am a URL shortener and you can use my services for free". Actually, this falls under the Meta-Information Request and Response protocols mentioned earlier.
- **Discoverability:** One desirable characteristic of a SM is the capability of discovering and connecting to other SMs dynamically. This specific characteristic of SMs would need a central repository to (a registry, part of the infrastructure of the ecology of social machines, not to be discussed in this paper as yet) register any SM that are likely to be discovered.

## 3.2. Classifying Social Machines

Out of many possible classifications, one way to look at social machines is through the simple taxonomy to the right, based on the types of interactions they have with each other, further described as follows:

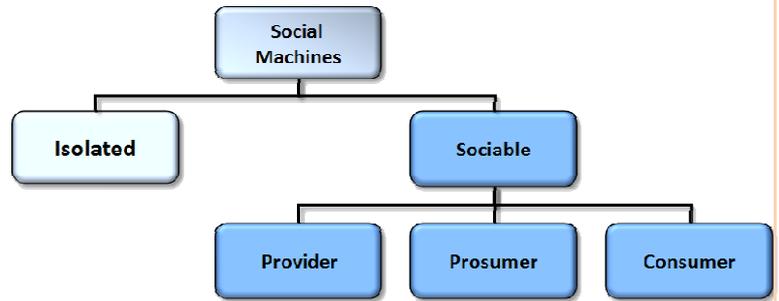

Figure 2: A taxonomy of Social Machines.

- **Isolated** – Social Machines that have no interaction with other Social Machines;
- **Provider** – Social Machines that provide services for other Social Machines to consume;
- **Consumer** – Social Machines that consume services that other Social Machines provide;
- **Prosumer** – Social Machines that both provide and consume services.

A Wrapper Interface of a Social Machine is the element responsible for the incoming and outgoing traffic of information and therefore an interface for the exchange services between Social Machines. If a given Social Machine does not provide means for interacting with its services then this Social Machine is classified as **isolated**. A typical example is a simple standalone application.

If a Social Machine has the ability to interact with other machines then it is said to be sociable. A Social Machine that provides services to other Social Machine can be classified as a **provider**. Applications such as Bitly, Twitter and Google Maps are all good examples of Social Machines as providers.

A Social Machine may also consume services that others provide. Web applications that consume services from other web applications (usually called mashups) are examples of Social Machines as **consumers**. Seesmic, Tweetdeck, HousingMaps and Wikipediavision consume services from Twitter, Google Maps and Wikipedia.

Finally, a Social Machine can both provide and consume services, being **prosumers**; they consume media and information from one or more Social Machines, process this information, combine it and provide the processed data to other Social Machines. The application described on our case study is an example of this type of Social Machine; Futweet is a SM that interacts with other applications and also provides its own services.

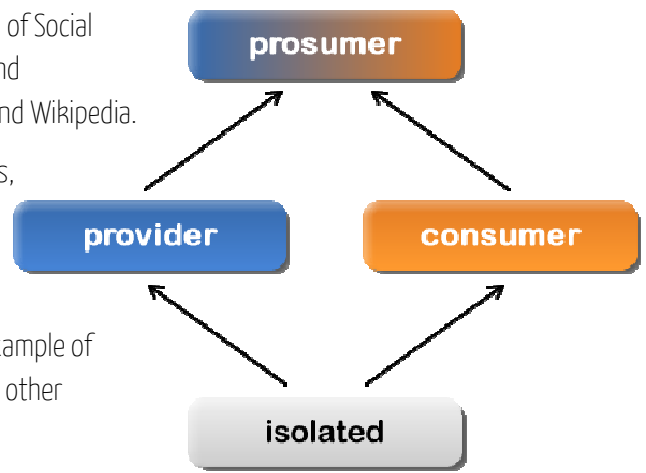

Social Machines as a partial order diagram.

One way to compare Social Machines is in terms of the complexity of the interactions between them, which is shown by the classification in **Figure 3**, that of a partially ordered set.

This set can be seen as a poset since certain pairs of its elements have a binary relationship which allows them to be arranged as predecessors and successors. **Figure 3** is also a Complete Lattice[2]; a lattice is a partially ordered set in which any two elements of the set have a common element that is greater than or equals to both elements (supremum), or lesser then or equals to both elements (infimum).

## 4. Case Study

In this section we describe a real system *(Futweet)* which was developed using the unifying idea we have been discussing so far. *Futweet* is both a social network and a guessing game about football (soccer) results. Initially developed for Twitter users, Futweet was subsequently connected with other online social networks, e.g. Facebook and Orkut, making it a good case study for illustrating the development of an application that uses the concept of Social Machine.

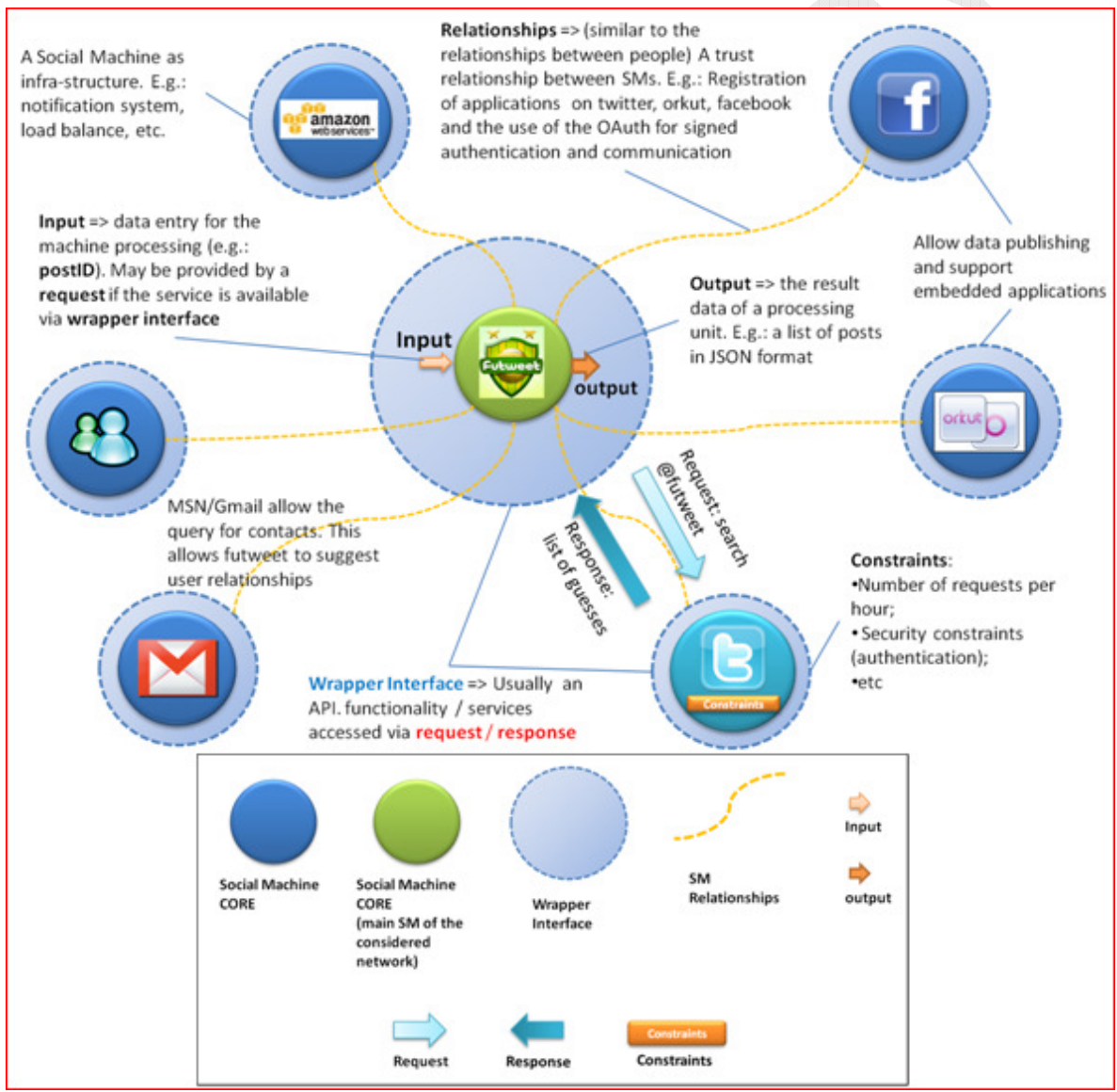

Figure 4: Futweet, a real network of Social Machines

---

[2] http://www.cis.upenn.edu/~cis610/discmath5.pdf

## 4.1. Futweet as a Social Machine

Futweet is a social game whose genesis comes from the idea of developing a social machine using the features provided by Twitter, which is a paradigmatic example of social machines. The game illustrates the development of a real social machine, since it was designed and built to be networked with other applications and be itself a connection point of other applications and services. In Figure 4, we show all the social machines tha comprise the Futweet social machine and their relationships.

*Futweet* is, of course, part of a network; in it, we can identify the main elements of a social machine, as discussed in Section 3:

- **Connections**: *Futweet* has connections (or relationships) with online social networks and other SMs (e.g.: Amazon EC2). These connections define components or services that can be considered part of the social game infrastructure. If any of these SMs are unavailable, *Futweet* as a whole may be affected.

- **Input and output:** Futweet uses the JSON format to input and output data from and to other SMs.

- **Processing Unit:** consists of *Futweet's* business rules in conjunction with the corresponding applications on the social networks *(Facebook* and *Orkut)*, and the mechanisms for interaction with Twitter, Gmail and MSN.

- **Requests and Responses**: they are managed by *Futweet's* API which is also responsible for the access of third party applications. Initially, Futweet's API is available only to implement its web interface.

- **Wrapper Interface**: It is represented by *Futweet's* API, which encapsulates the main features of the game available on the web.

- **Constraints:** Futweet has many constraints. Some of them are similar to the Twitter API rate limiting, e.g. number of records returned by the API (posts, guesses, users...). Futweet also limits request per account and IP.

- **States:** Information about the rate limiting of each service is usually presented inside the HTTP response header information. Futweet is also able to respond requests regarding specific meta information (e.g. Req – Give me *Futweet's* current state; Resp – Calculating rank, machine load is 99%).

The basic mechanism of the game involves the sending of guesses on soccer matches in a given league; such guesses are processed and compared with a set of pre-established scoring rules and the game winner is the user who gets more points at the end of a specified period, which generally coincides with the end of the championship in question. In the case of Twitter, sending the forecasts of a match follows a pre-defined syntax that has the team's acronyms and predicted scores. For instance, one may place a guess on a match by tweeting a string that obeys the following pattern:

> @futweet <TEAM1 Acronym> <Score for TEAM1> X <Score for TEAM2> <TEAM2 Acronym>

*Futweet* has an engine that periodically searches for tweets with this pattern, extracts the information that represents the guess of a user and then recalculates the overall rank. Since *Futweet* also exists as embedded applications on *Facebook* and *Orkut,* a user may request data (e.g. a ranking list) from *Futweet* the apps on top f those Social Machines. The case we have just described is an example of how Social Machines can work together to receive, compute and present information; *Futweet* is a social machine of class prosumer.

*Futweet* also has a relationship with *Amazon AWS*: it does not own the servers it runs upon and its infrastructure is provided in part by hosting its data on Twitter (guesses in the form of tweets) and by using the computational power provided by *Amazon EC2*. Thus, the social game is an application that is totally provided on the cloud, designed, implemented and available on the cloud. This reinforces the assertion that the fundamental component of a social machine (its processing unit) and its [possibly many] other components can be supported by other, existing, social machines, resulting in a network which is, by itself, the desired application.

Figure 4 provides an overview of Futweet as a social machine. The social game is presented as an architecture of related machines *(Twitter, Orkut, Facebook, Gmail and MSN)* working on top of a virtual infrastructure *(Amazon EC2)*. The connections with other social networks, Futweet's processing unit and the plataform provided by Amazon can also be considered a social machine by itself. The functionalities of this network are encapsulated by an API (wrapper interface) that makes the main features of the service available. It is important to note that the *Futweet* is the "glue code" between different social machines (with their own computational core).

## 4.2. Social Machine's Architecture Description

In order to better describe the elements and connections of a social machine-based architecture we are in the process of developing a Social Machine Architecture Description Language (SMADL) to specify social machine networks. SMADL's syntax is based on Armani's [Monroe, 2001], but we have added specific elements to support the concept of SMs and their relationships. The main elements of SMADL are represented by the meta-model depicted below.

All the elements defined in the social machine algebra (see Section 3) were mapped into SMADL constructors; SMADL provides the a number of core constructs for describing instances of Social Machine-based architectural designs:

- a **SocialMachineNetwork** is a collection of social machines, their relationships, and a description of the topology of a given social machine network.
- a **SocialMachine** is an entity representing the core computational units of a social machine network.
- **ProcessingUnit** represents the primary computational units of a social machine.
- **Input** is the data to be handled by the Social Machine's processing unit.
- **Output** is the result data returned by a Social Machine's processing unit after the execution of a core functionality.
- **States** correspond to the states that a social machine may assume.
- **Constraints** represent the restrictions over the behaviour of a social machine.

Paper still in α; new versions will be announced at twitter.com/srlm, #SocMac.   10

- **WrapperInterface** specifies the interface through which a SM externalizes its services and allows interactions with other SMs.

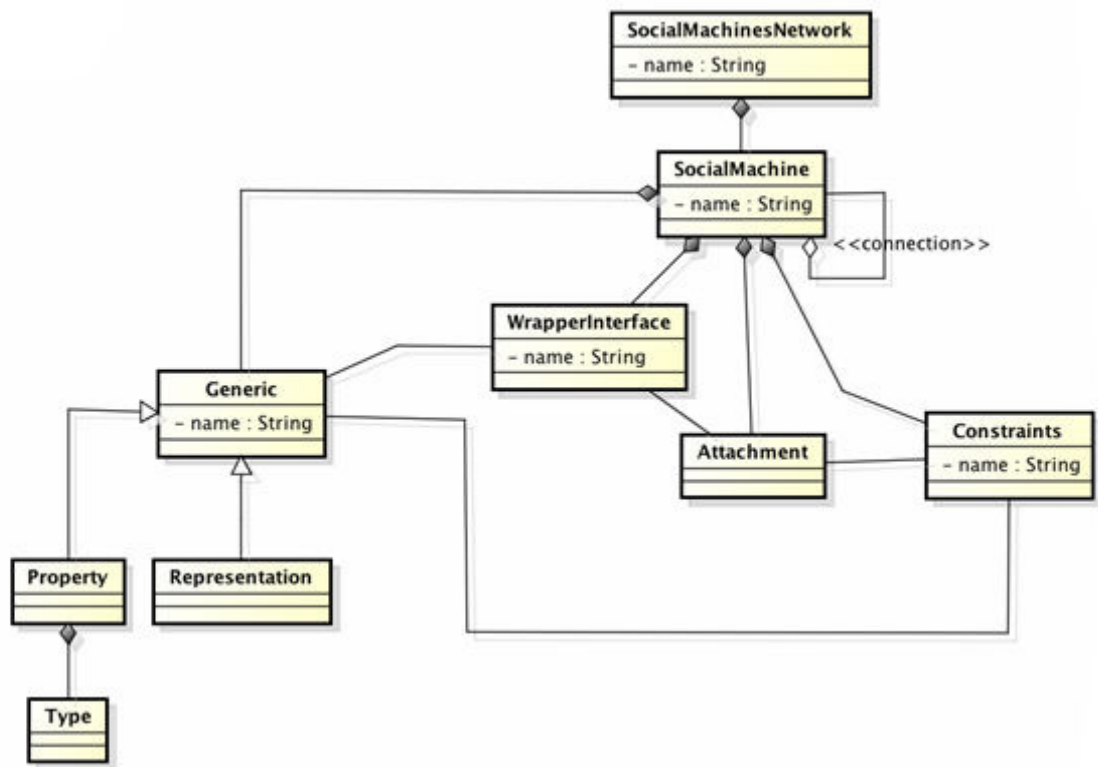

Figure 5: SMADL

- **Request** specifies the signature of a remote procedure call to a service defined by the SM's wrapper interface.
- **Parameters** are the parameters of a request.
- **Response** represents the reply to other SMs through the wapper interface.
- **Relationships** represent and mediate interactions between social machines.
- **ConnectionSettings** specify settings to be used in a relationship between two social machines.
- **Properties** are annotations that store additional information about other elements (*Constraints, Request, etc*).

The listing below summarizes the specification of the Futweet network in SMADL, with a lot of the details omitted to facilitate looking at the overall specification.

```
SocialMachineNetwork futweet_net = {

    SocialMachine twitter = {
        ProcessingUnit twitter_pu = {...};
        Constraint constraints = {...};
        WrapperInterface api = {...}}

    SocialMachine facebook = {...}
    SocialMachine amazon = {...}
    SocialMachine orkut = {...}
    SocialMachine gmail = {...}
    SocialMachine msn = {...}

    SocialMachine futweet_core = {

      ProcessingUnit fuweet_pu = {
          Input inputXml: xml;
          Output outputXml: xml;
          Input inputJson: json;
          Output outputJson: json;
          States {processing; idle; overload}};

      Constraint constraints = {Property request_per_hour < 5000;};

      WrapperInterface api = {

          Request doGuess = {
            Parameters {guesses:int [ ]};
            Response success: json;
            Property httpMethod="POST";
            Property url="http://futweet.com.br/futweet/palpitar";};

          Request getFutweets = {
            Parameters {filter:string};
            Response list: json;
            Property httpMethod="GET";
            Property url="http://futweet.com.br/futweet/getfutweets"}}}

      Relationships {
          (futweet_core to twitter) = {
              ConnectionSettings {name="Futweet";
                                  apikey:string;
                                  apisecret:string}};

          (futweet_core to facebook) = {...}}}
```

Listing 1: Futweet architecture in SMADL

## Project Considerations

During the design of Futweet as a social machine, it was necessary to consider a set of questions in which the answers had influences in the development of the social game:

- **Are there any available social machines on the web that could be (re)used by the project?** Building a web application as an SM should consider the existence of other machines to be (re)used. In the case of *Futweet*, already existing machines considered were: *Twitter, Amazon AWS, Gmail, MSN* and, thereafter, the online social networks *Facebook* and *Orkut*.

- **What is the utility that the social machine should provide for its environment (web)?** *Futweet* is one of many implementations of a soccer guessing game. However, its main goals are: *i)* providing mechanisms through APIs to allow users to use its platform to create their own applications of guessing game and *ii)* allowing the entertainment of Twitter users, extending the capabilities of Twitter through the addition of a new service.

- **What are the (read/write) operations provided by the application?** As seen in Section 3, social machines may have different social levels that vary according to *i)* the connections they have with other machines and *ii)* the type of operations that enable these relationships. *Futweet* is a prosumer social machine, it has connections to read/write on Twitter (read and put data in the social network) and allows the same operations through its own API (users can post and read data from *Futweet* remotely).

After having a set of (may be partial) answers for those questions, the implementation of *Futweet* was, in other words, designing a set of interfaces to access various social machines, governed by business rules (from the social game) that implied the functionalities and design of an API, on the top of which the application was also built. This simplistic view of *Futweet* is important for understanding the concept of social machines.

## 4.4. Implementation outline

The n-tier architecture described in Figure illustrates how the *Futweet* Social Machine **application** was designed and developed. Notice that the diagram does not specify any concept related to Social Machine. Indeed it is possible to describe a Social Machine using existing software engineering standards and tools. This is so because Social Machines is not a technology, a framework or a developing pattern, but a concept, a mental model for understanding and describing each and every entity connected to the web.

- **Client Tier**: A client application may be a browser, mobile app or even another web application. The request and response events are often HTTP, but they could de an email, XMPP messages, etc.

- **Interface Tier:** Within the application a request can be handled by a given web service that usually returns a result in –say- JSON or XML formats. But the client could request a page and, in this case, the web interface of the application would return HTML content and it could use Server Pages technologies to help deliver. There shouldn't be any business logic on this layer, only data validation and presentation.

- **Processing Tier:** After a request is validated by the interface tier it then may call services to process the data. This layer is responsible for business logic, algorithms and general processing, with helper libraries and classes also defined here.

- **Data Access Tier:** This layer is responsible for data access and persistence. DB connectivity is also defined and implemented here.

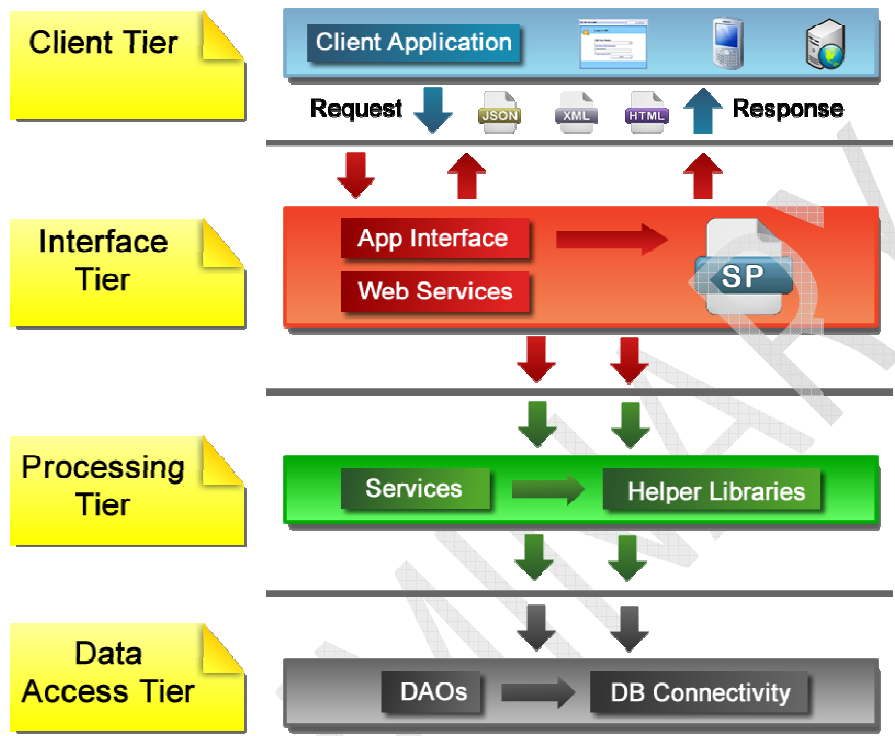

Figure 6. Futweet architecture diagram.

## 4.5. Considerations

There are several factors that should be taken into account when developing a social machine; one has to bear in mind that the complexity of a given system's development is directly related to the properties, power, limitations and restrictions of other social machines considered in the project. Non-functional requirements such as response time (in our project or design) can be affected by quality attributes of SMs being used as a basis for design and implementation, such as availability, limitations or restrictions of third party APIs, changes in the mechanisms for accessing social machines, among many other considerations.

The "emerging" in the title of this paper has a lot to do with that; for some time to come, it is likely that many people will resist to design applications in terms of social machines, especially when dependability is a serious matter, which is usually the case in (say) corporate systems. Even so, chances are that this market will develop quite fast when compared to (say) software reuse and software product line, given the huge gains in productivity that can be achieved here if we can depend upon stable, reliable social machines. Instead of reusing (for example) a piece of software that needs a supercomputer to run upon, one can just use (and not "reuse") the service provided by a supercomputer in the web, of course supported by that piece of software.

We have ho give in that Futweet is a "toy" project when we compare to the usual scale of corporate projects; but, considering the functionality made available by the effort and comparing that to what would have been the effort of developing it from scratch, we are not talking about a small project anymore; glue it is, of course, but it is glue that keeps together quite a lot of stuff provided by other, existing, parts of –Social Machines in- the web. This may be quite a novel way to develop and evolve corporate information systems, and a somewhat radical departure from the two current competing efforts, the old "all from scratch" and the difficult "reusing reusable parts".

While developing Futweet we have learned a few practical things that are likely to be part of the process of construction of many other social machines; a non-exhaustive list ought to include:

- **APIs – access limits**: some social machines are accessed by many application clients (other SMs); in order to survive such a barrage of requests, they typically have *rate limits* that establish restrictions related to the number of accesses to their APIs. Hence, it is important to analyze how the SM can be improved in order to decrease the number of accesses to other SMs.

- **Optimization related to the SM resources**: infrastructure platforms (ex.: Amazon EC2), have several mechanisms for charging the usage of hardware (bandwidth, processing and IO, among others). Thus, during the development of a SM which uses such type of infrastructure it is significant to identify more efficient and effective ways to use such resources.

- **Error propagation**: If your social machine is a network, whatever happens in parts of it outside your control will affect your performance and are likely to be identified by your users as being your problem; in Futweet, for example, problems in Twitter sometimes are interpreted by users as a problem of the Futweet system.

- **The social dynamics of machines may change over time:** It is important to note that the control of any changes related to social machines that are components of Futweet is external to the project itself. Thus, significant parts of the social game may function inconsistently due to changes in access mechanisms of the APIs or even (input) parameters that are sent in requests. During the development of *Futweet* Twitter changed some of its parts to avoid the Twitpocalypse [TwitP 2009] (which involves, among other factors, the depletion of identification numbers for the tweets) and this led to the malfunction of the whole game we identified the cause and patched it to fix the problems.

- **Absence of mechanisms for automatic verification of the status of services provided by social machines:** During the period of *Futweet*'s operation, especially championships with broad user base and large scale interaction as the World Cup, the lack of automated mechanisms for checking the availability of external services made it more difficult to monitor the operation of the game. In fact, in some cases there may be a long delay between the discovery of the instability of a given API and adaptation of the system to correct such glitches since the process is still largely manual.

- **Changes in the mechanisms in charging for API access:** Some major social machines are still grokking their business models, which is why today's free and public machines may become private and paid for later in time, which will of course affect the whole chain of the development.

- **Lack of mechanisms to ensure the quality of services provided by the social machines:** the relationships established between the social machines ought to be guaranteed by agreements upon quality of the service, response time or any other aspects impacting systems' performance. This being notoriously absent, in turn, in most current cases, reflects upon the quality of service provided by actual systems as a whole. We live in highly experimental times in the "web 3.0", where the complexity of building –say- subscription-based SMs as networks of other SMs is magnified by (still) some orders of magnitude. Even so, there are reasons to believe that there will be ways of contracting (say) SMs with pre-defined QoS soon in the future.

Last but not least, there is a severe problem of...

- **...Using specific tools and libraries**: Futweet was designed and built around three social networks, each of which having its own workings and set of rules; none of those was designed as a SM, we viewed them as SMs; that being the case, we had to (and for a time will have to) use software libraries made available by developers or, in cases, by the owners of such social machines to reduce the complexity of building Futweet. In our view, that is not going to be the case in the future, at least for commercial social machines; it is not only unnatural and counterintuitive, but also uneconomical to have to find and reuse code libraries to reuse services on the web. This being the case in the meanwhile, one has to take into account that the larger is the community of developers that make use of a particular social machine in their projects, the more interesting are the possibilities around that.

## 5. Conclusion and Future Developments – A Research Agenda

In social networks like Twitter and Facebook, there is no McLuhanian media as message; they are environments to form communities where each individual chooses to what and with whom to connect, to write their stories in a collective and shared way. In many aspects, people "are" the connections also, apart from being the content.

Rereading the history of the web, at the beginning there was an environment that we only could "read"; we saw and heard the network. We could ask something (to Google, for example) but that was not enough. There was nothing –or there was very little, in that early web–created by we ourselves. The emergence of platforms - which allowed the members of the audience, in the past, to become agents in the community, nowadays- has created a network where we can also "write." Writing, here, ranges from videos on YouTube to comments on blogs, or even connections we create and the stories we share on social networks.

The same goes for infrastructure to support business processes like salesforce.com; instead of programming computers, like we used to do, we will increasingly plan, develop and deploy our own web, uncovering horizons for innovation in depth, width, scale and speed, in a way that not even McLuhan or, for that matter, anyone else could have imagined even a few years ago.

The impact of this change will be revolutionary. Starting to program social machines, each one of us will be able to create her own applications and provide new forms of articulation and expression in the web. The emerging web of this decade, where we all will program, will be virtually equivalent to the Cambrian explosion in the past, where almost all animal phyla we came to know emerged. This networked explosion of ours will not happen by chance; it will be programmed in the form of Social Machines.

This is the key motivation that led us to define the very concept of Social Machines as a unifying way to define, describe and develop the complex and emerging web and its consequences, including whole new classes of web based information systems. To validate the idea we applied the concept of Social Machines while developing a real system, Futweet, described in this paper as a case study.

In order to see the big picture, we can frame in terms of the structure shown below (Figure 7) is as many nuber of layers as we wish, taking into account problems that vary from formalizing IaaS (Infrastructure as a Service) to defining metrics and processes for monitoring and guaranteeing them in a web of Social Machines.

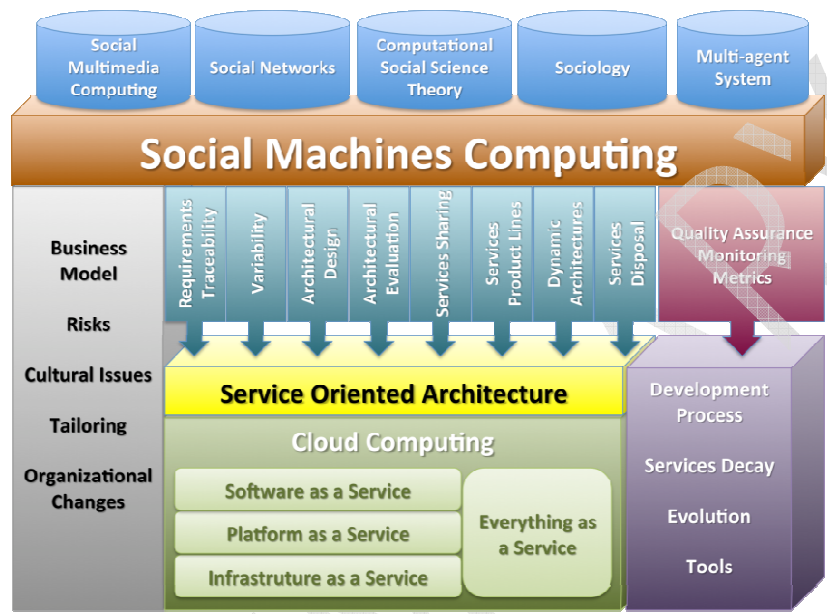

Figure 7. Social Machine Research Framework.

For future developments under this research framework, there is a large number of challenges which include (among many others):

- **Social Machines as a Unifying Architectural Framework for Defining and Developing Web-based Information Systems**: this is what this paper should be and what we aim at, given time (and competence);

- **SMADL as a Language to Specify Social Machines**: this includes language definition (syntax, semantics and pragmatics) and the development of methods, processes, tools and environments;

- **Ways, Formal and Less Formal, of Refining SMADL-defined Social Machines into Possible Implementations thereof**: the research question to be answered here is… is it possible to find simple ways to refine (transform) SMADL specifications into usable, real world implementations of SMs? That

- **A Variant of SMADL as a Language to Specify Contracts between Social Machines**: how do we define machine executable contracts between social machines? Would it be possible to write such contracts in a way that could empower independent (although constrained, for obvious reasons) negotiations between machines (i. e., could a machine ask for higher levels of QoS from another one, subject to limits of payment?…);

- **Conceptual, Implementation, Performance and Security Aspects of Social Machines**: SMs may represent particular challenges from one or more of such aspects; how would we deal with one or more of them? Which would be the general laws and rules for such?

- ... [Open: choose yours and suggest it to us; email silvio@meira.com] ...

- **A Privacy Policy Framework for Social Machine Computing**: this research project would aim to study the combined privacy challenges of social machines, trying to develop a common SM privacy policy framework based on open standards;

- **App Markets as Social Machines**: the goal here would be to study the challenges related to application markets at large (as they will be much more important than they are nowadays), treating them as special purpose SMs;

And, last but not least (and for completeness!)...

- **People as Social Machines**: the aim here would be to study the way people interact in and with the web and propose a an evolution of the concept of Social Machines described in this paper, considering people –peopleware and their inherently human capabilities- as an (old) new kind of Social (computing) Machine that can be used (like any other) in the development, deployment and evolution of applications on **the emerging web of social machines**.

# References [Partial]